\documentclass[acmtog]{acmart}

\usepackage{booktabs} 
\setcitestyle{square}
\usepackage{ifthen}
\usepackage{enumitem}
\usepackage{hyperref}
\usepackage{algorithm}
\usepackage[noend]{algpseudocode}
\usepackage{hyperxmp}
\usepackage{syntax}
\usepackage{amsfonts}
\usepackage{listings}
\usepackage{fancyvrb}
\usepackage{wrapfig}
\usepackage{graphicx}
\usepackage{subfigure}
\usepackage{xspace}
\usepackage{colortbl}
\usepackage{gensymb}
\usepackage{verbatim}
\usepackage{colortbl}
\usepackage{booktabs}
\usepackage{multirow}
\usepackage{cleveref}
\usepackage{pifont}
\usepackage{circledsteps}

\algdef{SE}[DOWHILE]{Do}{doWhile}{\algorithmicdo}[1]{\algorithmicwhile\ #1}%

\definecolor{mygreen}{rgb}{0,0.6,0}                         
\definecolor{mygray}{rgb}{0.95,0.95,0.95}
\definecolor{codebg}{rgb}{0.95, 0.95, 0.95}  
\usepackage{amsthm,amsmath,amssymb}
\usepackage{mathrsfs}
\usepackage[normalem]{ulem}

\renewcommand{\grammarlabel}[2]{#1\hfill#2} 
\definecolor{mypink}{rgb}{.99,.91,.95}

\begin{document}

\setcopyright{acmlicensed}
\acmJournal{TOG}

\title{DeepMill: Neural Accessibility Learning for Subtractive Manufacturing}





\author{Fanchao Zhong}
\authornote{Equal contribution}
\email{fanchaoz98@gmail.com}
\affiliation{%
  \institution{Shandong University}
  \city{Qingdao}
  \country{China}
}

\author{Yang Wang}
\authornotemark[1]
\email{1766897491wy@gmail.com}
\affiliation{%
  \institution{Shandong University}
  \city{Qingdao}
  \country{China}
}

\author{Peng-Shuai Wang}
\email{wangps@hotmail.com}
\affiliation{%
  \institution{Peking University}
  \city{Peking}
  \country{China}
}

\author{Lin Lu}
\email{lulin.linda@gmail.com}
\affiliation{%
  \institution{Shandong University}
  \city{Qingdao}
  \country{China}
}

\author{Haisen Zhao}
\authornote{corresponding author}
\email{haisenzhao@sdu.edu.cn}
\affiliation{%
  \institution{Shandong University}
  \city{Qingdao}
  \country{China}
}


%

\begin{abstract}
Manufacturability is vital for product design and production, with accessibility being a key element, especially in subtractive manufacturing. 
Traditional methods for geometric accessibility analysis are time-consuming and struggle with scalability, while existing deep learning approaches in manufacturability analysis often neglect geometric challenges in accessibility and are limited to specific model types.
In this paper, we introduce DeepMill, the first neural framework designed to accurately and efficiently predict inaccessible and occlusion regions under varying machining tool parameters, applicable to both CAD and freeform models.
To address the challenges posed by cutter collisions and the lack of extensive training datasets, we construct a cutter-aware dual-head octree-based convolutional neural network (O-CNN) and generate an inaccessible and occlusion regions analysis dataset with a variety of cutter sizes for network training.
Experiments demonstrate that DeepMill achieves 94.7\% accuracy in predicting inaccessible regions and 88.7\% accuracy in identifying occlusion regions, with an average processing time of 0.04 seconds for complex geometries.
Based on the outcomes, DeepMill implicitly captures both local and global geometric features, as well as the complex interactions between cutters and intricate 3D models.
\end{abstract}


\acmJournal{TOG}


\begin{teaserfigure}
\vspace{-2pt}
\centering
  \includegraphics[width=1.0\linewidth]{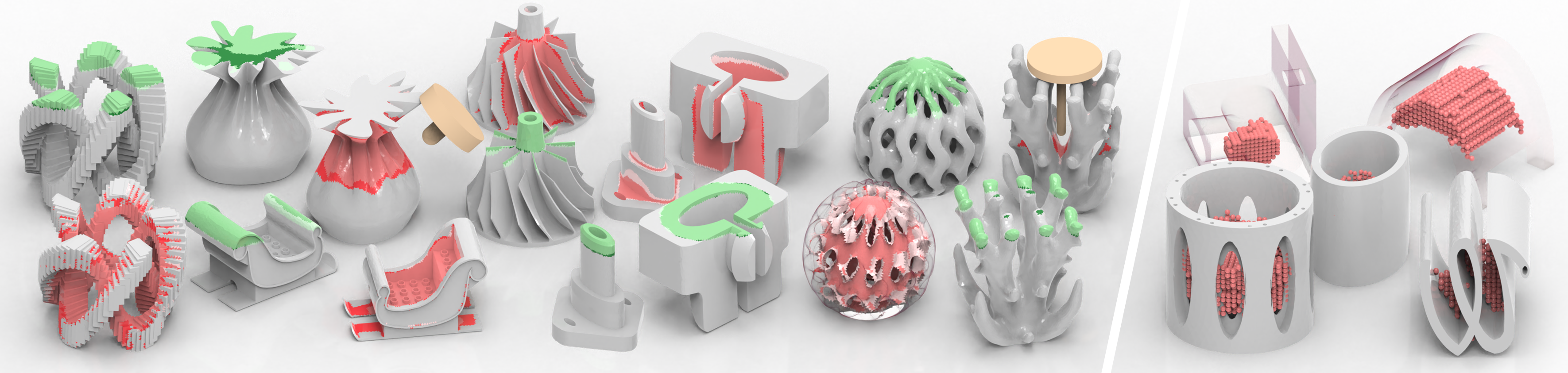}
	\vspace{-16pt}
\caption{We propose an octree-based neural network for cutter accessibility and severe occlusion detection on arbitrary meshes. Compared to traditional geometric methods, our network significantly reduces computation time, enabling real-time prediction during shape editing. Across various CAD and freeform model datasets with different mesh resolutions, the accuracy for inaccessible regions and occlusion regions reach 94.7$\%$ and 88.7$\%$, respectively. 
The cutter sizes for each shape in the figure are randomly generated. We display the cutter shapes used for two shapes. On the left, the red and green areas represent inaccessible and severely occlusion regions predicted by the network. Dark red (green) and light red (green) areas indicate under- and over-predictions, respectively, compared to the traditional geometric method. On the right, the results of applying our network for volume accessibility analysis are shown. }
  \label{fig:teaser}    
\end{teaserfigure}

\newcommand{\hsyntax}[1]{\ensuremath{\mathit{#1}}}
\renewcommand{\grammarlabel}[2]{#1\hfill#2}

\newcommand{\manuAna}{{manufacturability analysis}\xspace}
\newcommand{\accessAna}{{accessibility analysis}\xspace}


\maketitle

\section{Introduction}
\label{sec:intro}

\begin{teaserfigure}
\vspace{-2pt}
\centering
  \includegraphics[width=1.0\linewidth]{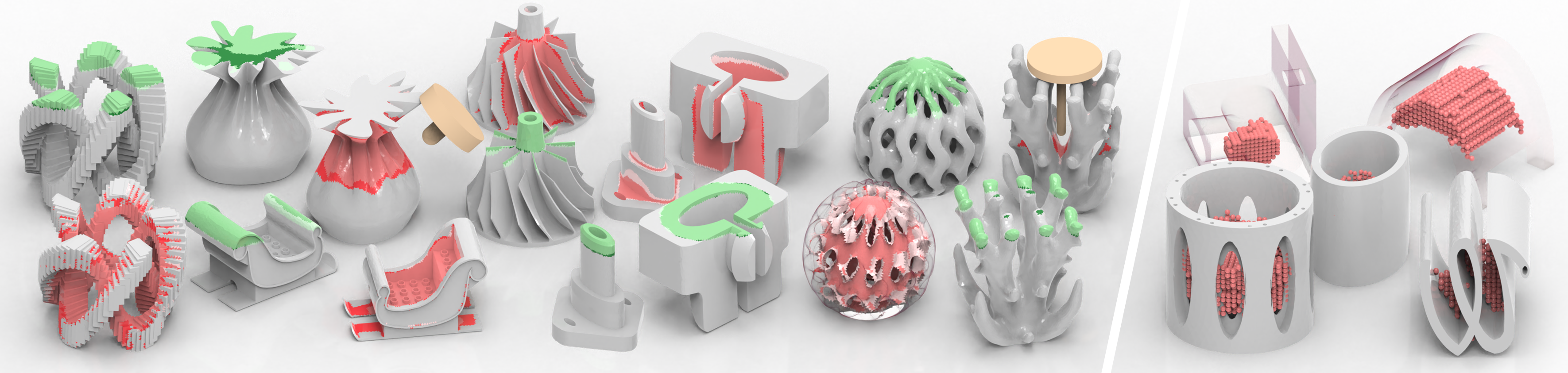}
	\vspace{-16pt}
\caption{We propose an octree-based neural network for cutter accessibility and severe occlusion detection on arbitrary meshes. Compared to traditional geometric methods, our network significantly reduces computation time, enabling real-time prediction during shape editing. Across various CAD and freeform model datasets with different mesh resolutions, the accuracy for inaccessible regions and occlusion regions reach 94.7$\%$ and 88.7$\%$, respectively. 
The cutter sizes for each shape in the figure are randomly generated. We display the cutter shapes used for two shapes. On the left, the red and green areas represent inaccessible and severely occlusion regions predicted by the network. Dark red (green) and light red (green) areas indicate under- and over-predictions, respectively, compared to the traditional geometric method. On the right, the results of applying our network for volume accessibility analysis are shown. }
  \label{fig:teaser}    
\end{teaserfigure}

Manufacturability is a fundamental concept in product design and production~\cite{shukor2009manufacturability,gupta1997automated}, referring to the ease and efficiency with which a design can be transformed into a physical product, while accounting for factors such as material constraints, cutter capabilities, and production costs~\cite{joshi1988graph,li2006machinability,hoefer2017automated}.
Ensuring manufacturability early in the design process is critical to avoiding costly revisions, delays, and inefficiencies, thereby optimizing production timelines and minimizing resource waste.

Accessibility is a key aspect of manufacturability~\cite{elber1994accessibility,spyridi1990accessibility}.
In subtractive manufacturing, accessibility pertains to whether all surfaces and features of a part can be reached by machining tools during production~\cite{zhang2020manufacturability}. Accessibility issues arise when certain features are difficult to access, such as deep holes, internal cavities, or overhanging geometries. Addressing these challenges through accessibility analysis is essential for identifying potential machining difficulties early, enabling designers to adjust part geometry or select appropriate cutters to optimize the manufacturing process. Beyond subtractive manufacturing, \accessAna also plays a critical role in decisions such as setup planning~\cite{zhong2023vasco}, cutter selection~\cite{athawale2010topsis}, cutter orientation adjustment~\cite{mahdavi2020vdac}, and tool path planning~\cite{balasubramaniam2003collision}, contributing to overall production efficiency and cost-effectiveness.

Traditional methods for geometric accessibility analysis, which emerged in the 1990s, primarily rely on geometric and computational techniques to evaluate cutter accessibility in multi-axis CNC machining. These methods, while foundational, are often time-consuming, especially for geometrically complex parts, with analysis of intricate designs taking hours—unacceptable in fast-paced design environments that require rapid iteration~\cite{dai2018support}. 
Although early approaches advanced from basic visibility analysis to more precise accessibility evaluations, they struggle with scalability and high computational overhead when applied to high-resolution 3D models. These limitations highlight the need for faster, more scalable methods capable of handling complex geometries and diverse cutter parameters.

In recent years, the advent of deep learning techniques has opened new possibilities for improving computational efficiency in manufacturability analysis.
Several studies have explored the use of deep learning models to predict non-manufacturable regions~\cite{kerbrat2011new,chen2020manufacturability,ghadai2018learning}, enhancing performance by reducing processing time. 
However, most of these efforts focus primarily on process-related issues, such as process planning and collision detection~\cite{chen2020manufacturability}, while neglecting the crucial geometric challenges inherent in accessibility analysis.
Moreover, these methods often rely on feature-based CAD models~\cite{yan2023automated,balu2020orthogonal}, which limits their applicability to freeform or highly complex product designs.

This paper presents DeepMill, the first neural framework, to the best of our knowledge, specifically designed for predicting non-manufacturable regions in arbitrary models, including freeform shapes, with high accuracy and efficiency. Unlike previous approaches, DeepMill focuses specifically on cutter accessibility, identifying regions where cutter collisions occur due to geometric constraints such as occlusion. 
We propose to utilize a neural network capable of real-time predictions for both non-manufacturable regions and the occlusion regions causing these issues, providing designers with actionable insights to quickly iterate and refine their designs. 
DeepMill demonstrates exceptional generalization across various cutter sizes and complex geometries, making it suitable for a wide range of design contexts.

One of the key challenges in accessibility analysis is the complexity of cutter collisions, which can be both local and global in nature. Factors such as cutter rotation and size affect accessibility, requiring methods that efficiently learn and represent these underlying geometric features. 
Moreover, the scarcity of extensive training datasets for these specific tasks has hindered the creation of robust models.

To overcome these challenges, we propose utilizing octree-based convolutional neural network (O-CNN) to efficiently capture both local and global geometric features, while embedding cutter modules to capture intricate interactions between cutters and complex 3D surfaces.
This approach enables our network to handle both CAD and freeform models, providing a scalable and flexible solution to the manufacturability analysis problem.

Additionally, we created the first inaccessible and occlusion regions analysis dataset with diverse cutter parameters for training DeepMill and generated multiple test set categories, addressing the challenges of data scarcity.

In summary, DeepMill offers a significant advancement in both computational efficiency and accuracy. 
Experiments indicate DeepMill achieves 94.7$\%$ and 88.7$\%$ accuracy on average in identifying inaccessible and occlusion regions, with an average processing time of only 0.04 seconds for complex geometries.
Our model is adaptable to a wide range of cutter sizes, ensuring its applicability across diverse design contexts. 
Additionally, we introduce a new dataset to support further research in this area and facilitate the development of more robust manufacturability analysis cutters.

\section{Related Work}
\label{sec:related}

The rapid advancements in artificial intelligence (AI) have significantly propelled solutions in digital geometric design and manufacturing~\cite{abdelaal2024ai}. 
AI applications in additive manufacturing~\cite{wang2020smart, zhang2024machine} and subtractive manufacturing~\cite{manikanta2024machine, soori2023machine} have been extensively reviewed.
Cutting-edge research similarly advances computer graphics, encompassing assembly planning~\cite{jones2021automate}, LLM-centric design and manufacturing~\cite{makatura2024can1,makatura2024can2}, 3D printing path optimization~\cite{huang2024learning, liu2024neural}, and feedback-based 3D printing control~\cite{piovarci2022closed}.
These advancements underscore a significant trend towards AI-based automation and optimization in design and manufacturing processes.
This paper primarily examines the \manuAna using a learning method, specifically focusing on the \accessAna for subtractive manufacturing~\cite{gupta1997automated, hoefer2017automated}, which is also the main content of this section.

\paragraph{Traditional Manufacturability Analysis}


Manufacturability of subtractive manufacturing is defined as four characteristics: visibility, reachability, accessibility, and setup complexity \cite{gupta1997automated, hoefer2017automated}.
Traditional techniques of \manuAna primarily relies on two approaches: feature-based and feature-less methods~\cite{zhang2020manufacturability}. Feature-based methods extract machining features as a prerequisite, using techniques like graph-based analysis~\cite{joshi1988graph}, volumetric decomposition~\cite{kailash2001volume,kim1990convex}, and hint-based approaches~\cite{regli1995geometric}. Feature-less methods, on the other hand, analyze surface representations to assess manufacturability, employing techniques such as slicing for machinable range mapping~\cite{li2006machinability} and octree decomposition~\cite{kerbrat2011new}.


Aligned with these conventional studies, where accessibility is the main evaluation metric for both feature-rich and feature-agnostic methods~\cite{zhang2020manufacturability}, this paper primarily focuses on assessing accessibility.
Unlike traditional approaches, we investigate the use of neural networks for \accessAna.

\paragraph{Learning based Manufacturability Analysis}

Early efforts to integrate machine learning into \manuAna have demonstrated potential in addressing the limitations of traditional methods. Examples include heuristic rule-based scoring~\cite{kerbrat2011new,joshi2017geometric} and process planning for Quasi-rotational components~\cite{chen2020manufacturability}. Recent advancements leverage deep learning, such as autoencoder-based generative models for feature matching~\cite{yan2023automated}, 3D-CNNs with orthogonal distance fields (ODF) for manufacturability prediction~\cite{balu2020orthogonal}, and enhanced B-rep structures with surface normal data for feature recognition~\cite{ghadai2018learning}. Hierarchical graph neural networks have also been applied to analyze B-rep topology and UV network geometry for multi-level learning~\cite{huang5065158hierarchical}.

A significant limitation of the above studies lies in the challenge of applying learning methods to more intricate aspects, such as predicting cutter accessibility for freeform shapes.
To the best of our knowledge, no prior research has introduced a learning-based \accessAna method applicable to freeform and complex geometries. Our work bridges this gap by enabling real-time prediction of non-reachable and occlusion areas while supporting general cutting tools.

\paragraph{Geometric Accessibility Analysis}


Determining accessibility in multi-axis CNC machining remains a significant challenge, with numerous methods proposed since the 1990s. Early research focused on geometric and computational approaches, with Woo’s spherical visibility map laying the foundation for 3D surface visibility analysis~\cite{woo1994visibility,elber1994accessibility}, and Spyridi and Requicha introducing global and local accessibility concepts for CMM~\cite{spyridi1990accessibility}. Other methods explored NURBS surfaces for interference calculations~\cite{lee19952}, configuration space mapping for cutter range feasibility~\cite{choi1997c}, and effective cutter radius evaluations for end mills~\cite{vafaeesefa1998accessibility}.
To simplify calculations, methods such as sampling-based orientation analysis~\cite{dhaliwal2003algorithms,zhao2018dscarver,mahdavi2020vdac}, plane projections for blisk machining~\cite{chen2015collision}, and boundary-based range construction~\cite{liang2016accessible} have been widely adopted. More advanced methods include Gaussian spherical mapping~\cite{liu2020sequence} and ray-tracing metrics for assessing five-axis milling manufacturability~\cite{chen2021design}.

These approaches reflect the progression from basic visibility analysis to more efficient and precise accessibility evaluations in complex machining scenarios. However, when applied to intricate or high-resolution 3D models, they often face scalability challenges and significant computational overhead due to the time-intensive nature of strictly geometric methods~\cite{dai2018support}. Therefore, developing fast and scalable methods capable of handling complex geometries and diverse cutter parameters is highly meaningful.

\paragraph{Spatial Analysis Learning}

Early works extended deep learning methods to 3D voxels~\cite{Maturana2015,Wu2015} for spatial analysis.
However, voxel-based approaches are computationally expensive and memory-intensive, making them unsuitable for high-resolution 3D data.
To address these limitations, sparse voxel-based CNNs leverage octrees~\cite{Wang2017} or hash tables~\cite{Graham2018,Choy2019} to confine computation to sparsely occupied voxels, significantly improving efficiency.
Point-based neural networks~\cite{Qi2017a,Qi2017,Li2018} eliminate the need for voxelization by directly processing point clouds, offering an alternative solution.
Recently, transformers have also been applied to 3D data, demonstrating promising results~\cite{Guo2021,Zhao2021,Wang2023}.
In this work, we adopt O-CNN~\cite{Wang2017} for \accessAna due to its efficiency and strong performance across a range of 3D tasks.

\section{METHODOLOGY}
This section first outlines the problem and goal of the network, followed by an introduction to our octree-based network, DeepMill, designed for predicting inaccessible and occlusion regions on the surface of the input mesh $M$. We then provide a detailed analysis of its advantages over GNN-based methods.

\subsection{Problem formulation}
The detection of inaccessibility and occlusion regions, which segments these areas from the surface of $M$, is formulated as a 3D geometric segmentation problem. 
To facilitate the computation, we use points to represent its local region.
The goal of inaccessibility detection is to identify inaccessible points (labeled as $l_I$) on $M$ where the cutter cannot reach without collision, while occlusion detection aims to locate points causing the most severe occlusion for the inaccessible points. 
In this paper, we classify the top 10$\%$ of points with the highest occlusion severity as "occlusion points" (labeled as $l_O$).
Due to the severe imbalance in data distribution, we use the F1-score to evaluate the occlusion points.
Unlike standard visibility problems, the shape and size of the cutter $\mathcal{C}$ directly influence the segmentation results. Consequently, this task is formulated as a dual-task binary segmentation problem with a cutter-aware objective, expressed as:
\begin{equation}
\label{eq:weightSet}
\begin{aligned}
Max(accuracy(l_I)+F1(l_O)\;|\;\mathcal{C}).
\end{aligned}
\end{equation}

\begin{figure}[t]
\centering
\includegraphics[width=1.0\linewidth]{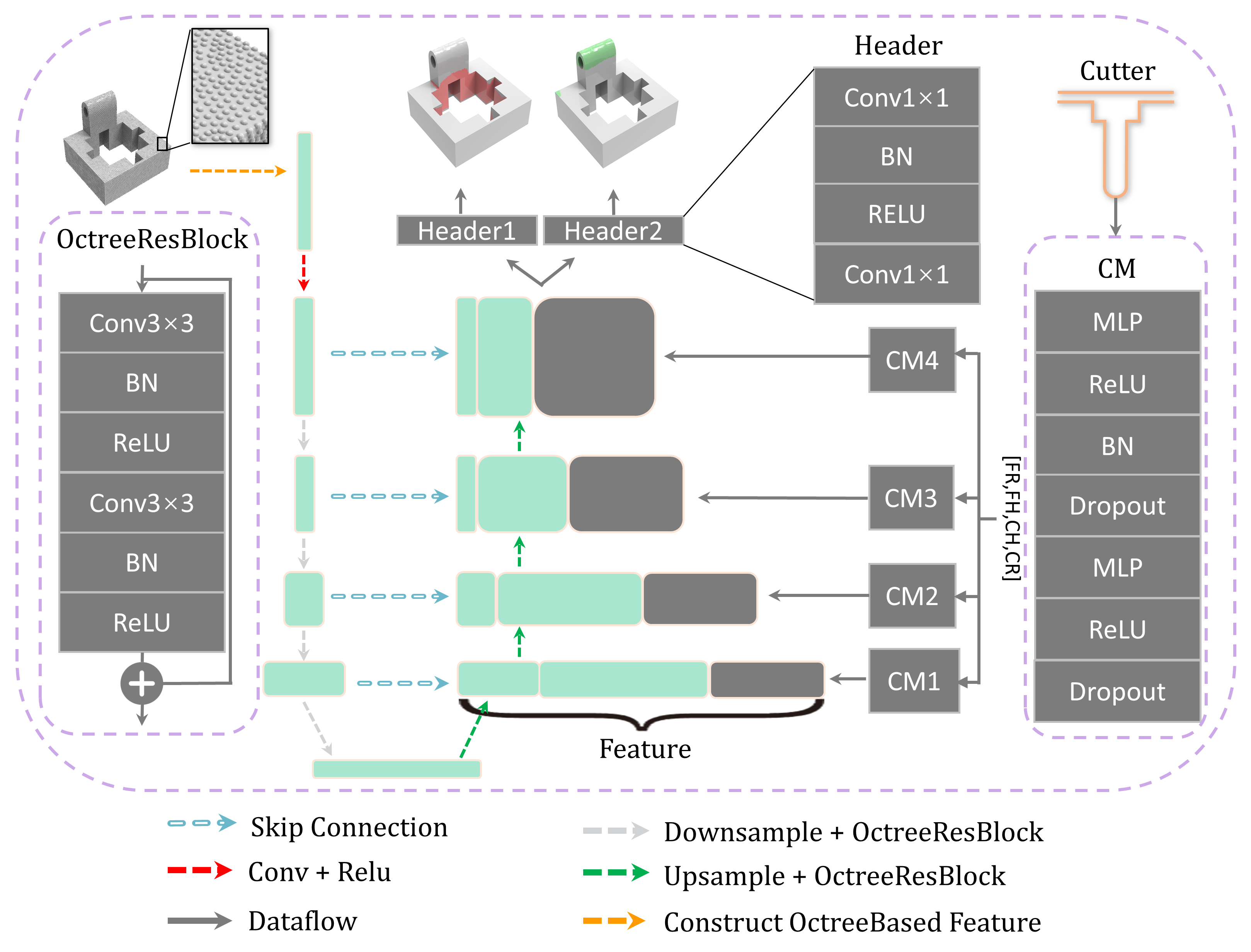}
\caption{DeepMill's network architecture. The input mesh is converted into a point cloud with normals, with each point corresponding to a Voronoi cell's site. Features are progressively extracted through the encoder, and the decoder, embedded with a cutter module (CM), restores spatial resolution. Both encoder and decoder are stacked with several Octree-based residual blocks. Finally, each site is subjected to dual-task binary segmentation through two header layers. Red and green represent the inaccessible and occlusion regions, respectively.}
\label{fig:Cutter-embended}    
\end{figure}

\subsection{DeepMill}
DeepMill's main components include the encoder and the decoder, which are well-suited for segmentation task. Additional cutter modules and prediction head are added to further adapt to our problem. \autoref{fig:Cutter-embended} shows the network architecture of DeepMill.

\paragraph{O-CNN with U-Net architecture.}
Unlike more complex network architectures, such as the Hierarchical Graph Neural Network~\cite{huang5065158hierarchical}, the O-CNN-based U-Net architecture uses a concise representation—point clouds and their normals—as input. The output consists of two predicted labels for each point. 
The U-Net architecture~\cite{ronneberger2015u} is composed of an encoder and a decoder, both of which are stacked with multiple Octree-based residual blocks~\cite{wang2017cnn}, with skip connections between the encoder and decoder. 
The encoder progressively extracts multi-scale features from the input 3D data, while the decoder gradually restores spatial resolution and reconstructs accurate predictions through the use of skip connections. 
DeepMill adopts this architecture, and the benefits of this approach will be discussed in \autoref{sec:analysis}.

\paragraph{Cutter embeding}
To enable the network to learn the impact of the cutter on inaccessibility and occlusion points, we embed cutter modules into the network. Compared to the encoder, the decoder is closer to the network's final decision-making region, and concatenating the cutter features at this stage minimizes interference with the network's early learning of geometric shapes. We validate this hypothesis in subsequent experiments. Furthermore, considering that the cutter causes collisions in both local regions and distant global regions (collide with the shaft space above the cutter) of $M$, we embed cutter modules at every layer of the decoder to help the network better learn the collision patterns between the cutter and $M$ at different scales.

In detail, we encode the four shape parameters $\{CR, CH, FR, FH\}$ of the cutter into a vector $\mathbf{V} = [v_1, v_2, v_3, v_4]^T$ and pass it through four fully connected cutter modules. As shown on the right side of \autoref{fig:Cutter-embended}, each cutter module consists of two 'Linear-ReLU-BN-Dropout' sub-blocks, where the 4-dimensional vector $\mathbf{V}$ is transformed into a 256-dimensional cutter feature. These features are then concatenated into each layer of the decoder in the U-Net architecture.
\begin{equation} 
f'_i = f_i \oplus f_i^c, \quad i = 1, 2, 3, 4
\end{equation}
where $f_i$ indicates the output feature of the $i$-th layers in the decoder and $f_i^c$ is the output feature of the $i$-th cutter module.

\paragraph{Dual-head segmentation} 
To predict inaccessible points and occlusion points separately, we use two fully connected header layers to predict these two types of labels. Since occlusion points are calculated based on inaccessible points, and both labels are computed using the same geometric algorithm during collision detection, there is a strong geometric correlation between the two labels. Therefore, before passing through the header layers, their features are fully shared. The predicted results $\hat{y}^i_j$ for $s_i \in S$ are denoted as:
\begin{equation} 
\hat{y}^i_j = \text{header}_j(f'_4), j=1,2
\end{equation}

\paragraph{Architecture details} 
As shown on the left side of \autoref{fig:Cutter-embended}, each octree-based residual block consists of two 'Convolution + BN + ReLU' sub-blocks, connected by residual connections~\cite{he2016deep}. Batch normalization (BN) is applied to reduce internal covariate shift~\cite{ioffe2015batch}, while the ReLU activation function $(f : x \in \mathbb{R} \longmapsto max(0,x))$ is used to activate the output.

In the encoder, the input point cloud undergoes multiple octree-based 3D convolution operations through several octree-based residual blocks, generating feature maps at different levels to capture multi-scale geometric features for hierarchical representation. Unlike traditional 3D-CNN convolutions~\cite{maturana2015voxnet}, the octree structure marks non-empty nodes at the current depth, representing regions containing point clouds, and applies convolutions only to these nodes. 
The depth of the octree gradually decreases, and high-resolution child node features are aggregated into their corresponding parent nodes.

In the decoder, the global feature map is progressively processed through deconvolution for feature upsampling and spatial detail recovery, with cutter feature fusion enhancing the modeling of inaccessibility and occlusion effects.
As the depth of octree increases, features are progressively passed down to the high-resolution child nodes.
Output-guided skip connections~\cite{wang2020deep} are used to transfer features from the encoder to the decoder, excluding sparse regions. If the octree node corresponding to a feature output from a block is empty, the skip connection is not applied.

\paragraph{Loss function}
During network optimization, we use cross-entropy loss function to compute the loss for inaccessible and occlusion points separately, denoted as $\mathcal{L}_I$ and $\mathcal{L}_O$. The total loss function is defined as:
\begin{equation}
\mathcal{L} =  \mathcal{L}_I(\hat{y}_1, y_1) +  \mathcal{L}_O(\hat{y}_2, y_2)
\end{equation}
where $y_1$ and $y_2$ denote the ground truth labels calculated using the geometric method mentioned in \autoref{sec:Overview}.

\begin{figure}[t]
\centering
  \includegraphics[width=1.0\linewidth]{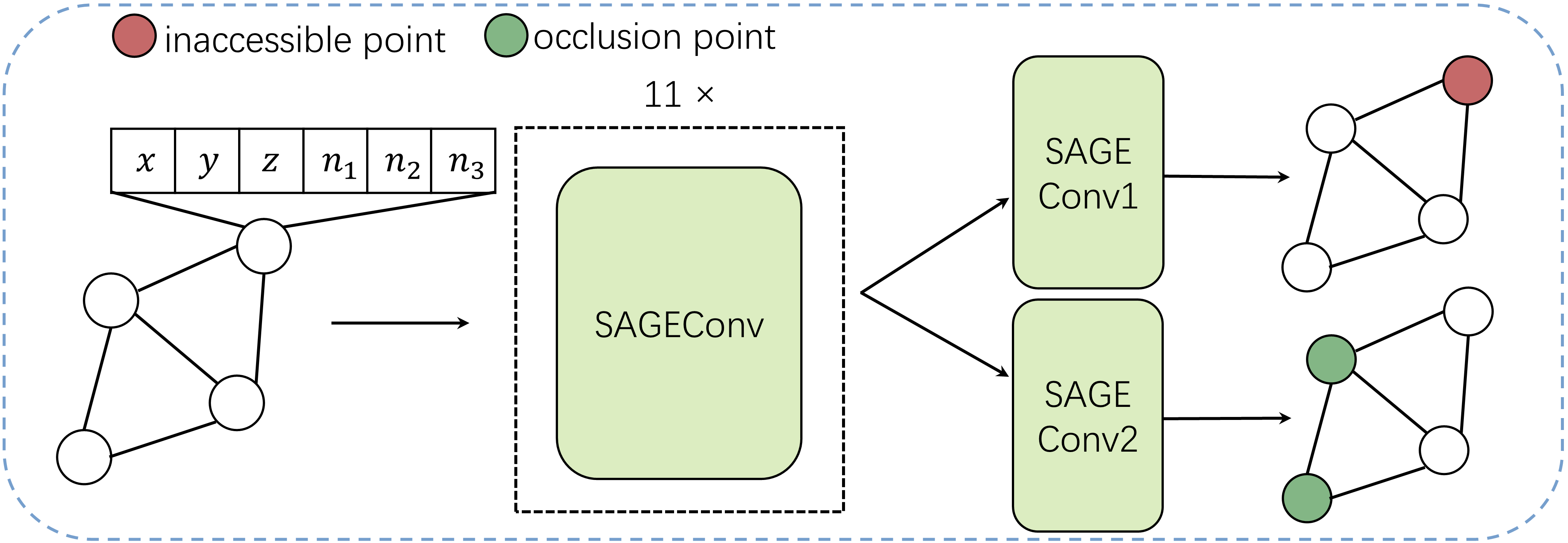}
\caption{Comparisons with GraphSAGE. $M$ is converted into a graph, with nodes representing mesh vertices and edges representing topological connections of them. Similar to O-CNN, initial node features include vertex coordinates and normals. Node features are propagated and updated through successive convolutions on neighboring nodes. }
  \label{fig:different-methods}    
\end{figure}

\subsection{Analysis for different networks}
\label{sec:analysis}
To demonstrate the simplicity and effectiveness of DeepMill, we constructed GNN-based framework similar to the approach in \cite{huang5065158hierarchical} for comparison. As shown in \autoref{fig:different-methods}, we used the classic GraphSAGE model \cite{hamilton2017inductive}. 
The network generates node embeddings by sampling neighboring nodes and aggregating their features, addressing the computational bottleneck of traditional GNNs on large-scale graphs. Compared to GCN~\cite{kipf2016semi} or GAT~\cite{velivckovic2017graph}, it is better suited for handling graphs constructed from high-resolution meshes in accessibility analysis.
GraphSAGE seems capable of preserving the mesh's topology, thus learning the geometric relationship between the cutter and $M$ more effectively. However, there are significant global collisions between the cutter and $M$, and occlusion points are often topologically distant from inaccessible points, making it difficult for graph convolutions to efficiently capture this relationship. 

In contrast, O-CNN-based U-Net architecture efficiently processes sparse 3D data through the octree structure, avoiding redundant computations. In accessibility analysis, CAD shapes often contain numerous holes, grooves, etc., and O-CNN effectively captures these key sparse geometric features. 
Additionally, the multi-scale convolution operation based on 3D space extracts shape features at different scales and is more effective at capturing collision relationships between distant positions on the mesh.
Furthermore, U-Net’s encoder-decoder structure with skip connections enables it to capture both local features and global context, preserving the detailed geometric insights essential for predicting inaccessible points.

\section{Geometric method for dataset generation}
\label{sec:Overview}
In this section, we introduce a rapid geometric approach to generate datasets with labels for inaccessible and occlusion regions.

\begin{figure}[t]

\centering
\includegraphics[width=1.0\linewidth]{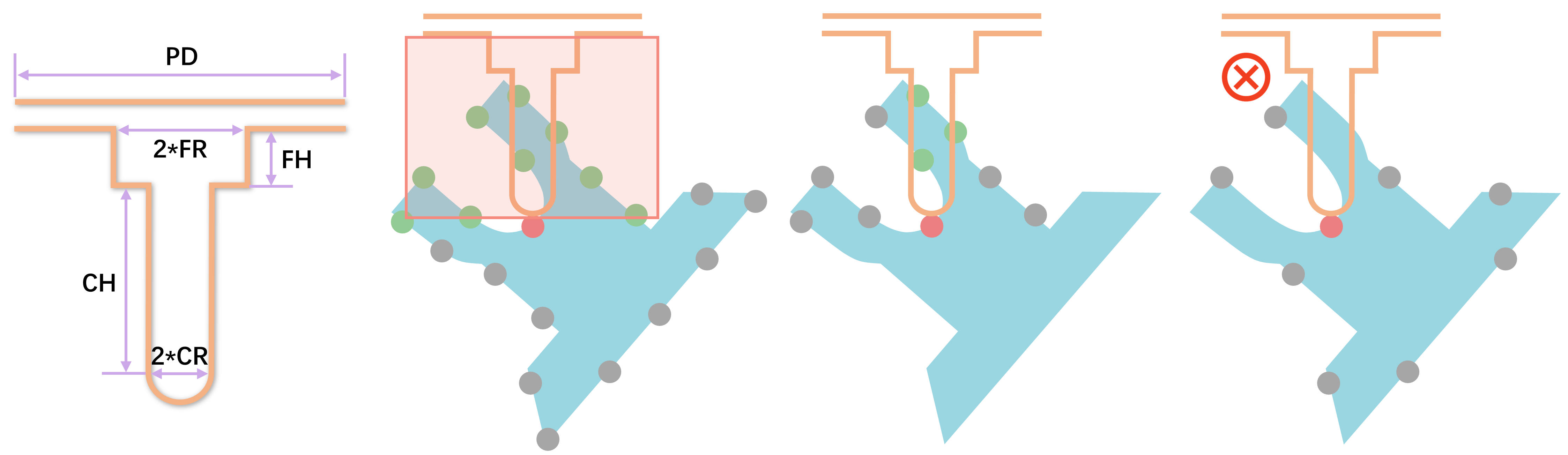}
\leftline{ \footnotesize  \hspace{0.08\linewidth}
            (a)  \hspace{0.21\linewidth}
            (b)  \hspace{0.22\linewidth}
            (c)  \hspace{0.21\linewidth}
            (d)}
    \caption{Illustration of inaccessible point detection. Orange represents the cutter, and the gray points represent sampled Voronoi sites. (a) A ball-end cutter can be simplified using four parameters. Note that above cutter is a non-accessible shaft space, and $PD$ is set to infinity. (b) For collision detection with red points, the mesh is first rotated, and points are quickly filtered by checking whether they lie within the detection box (red) of radius FR+$\sigma$, which eliminates most points far from the cutter. $\sigma$ is set to 5 in our experiments. (c) A finer collision check is performed for the points inside the box. (d) To prevent the cutter from penetrating the mesh without detection, the spacing between adjacent sites must be smaller than the cutter’s ball-end radius ($CR$).}
\label{fig:collision-detection}    
\end{figure}

\subsection{Voronoi-based accessibility analysis}
\label{sec:acc}
We use the subtractive collision detection method from \cite{zhong2023vasco} to gather accessibility training data, as it's efficient. 
We introduced a slight modification to their method by incorporating a detection box for pre-detection, enabling faster calculation.
Details of the method are outlined below.

\paragraph{Voronoi-based sampling.} The inset figure demonstrates the use of \begin{wrapfigure}[6]{r}{0.18\textwidth}
\vspace{-14pt}
\hspace{-20pt}
\centering
\includegraphics[width=0.19\textwidth]{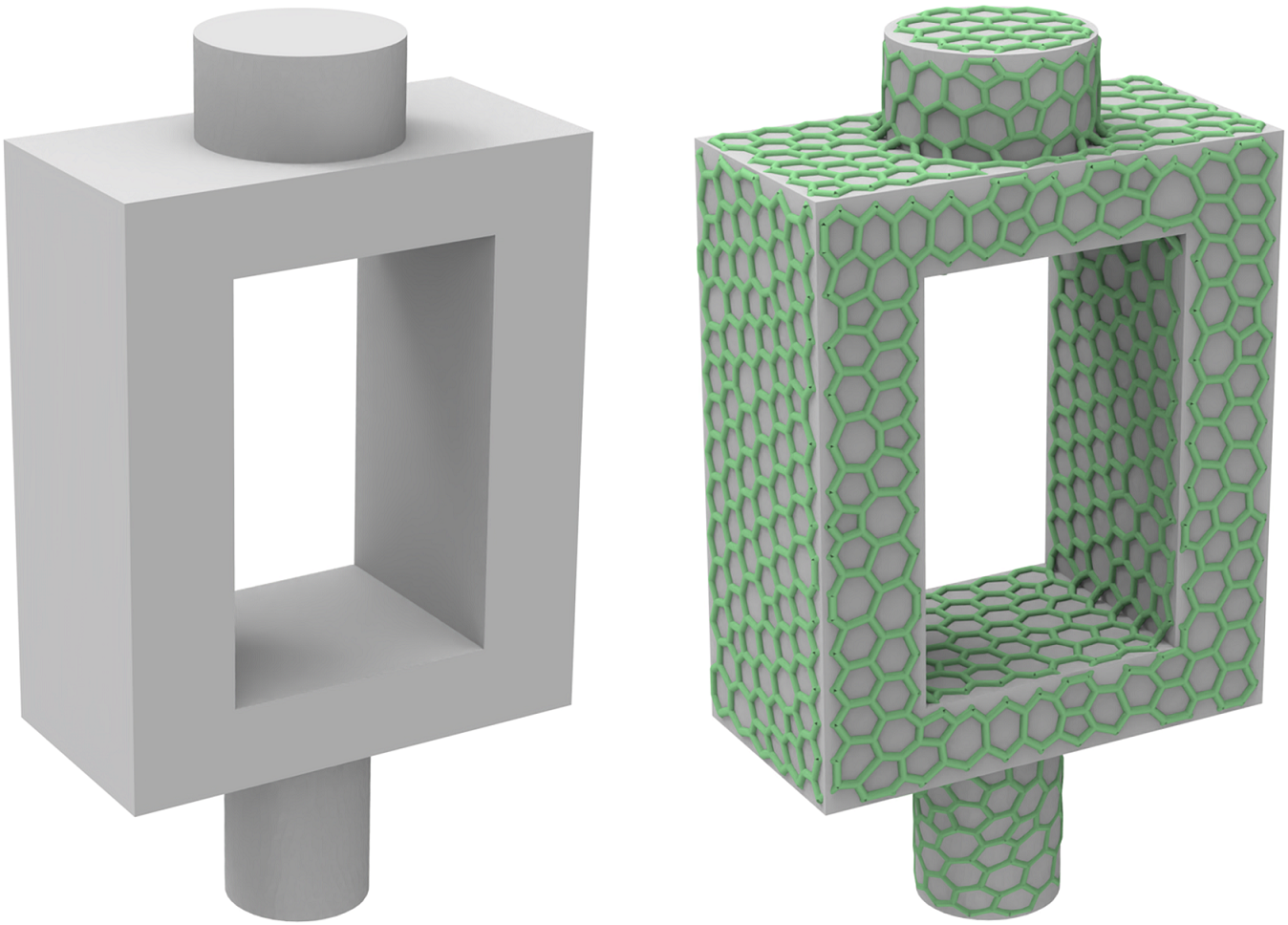}
\end{wrapfigure} Lloyd's Voronoi relaxation for uniform sampling on the surface of $M$, with each Voronoi cell represented by its site $s_i\in S$, where $S = \{ s_1, s_2, \dots, s_n \}$. The surface of $M$ can be simplified as $M = \cup_{i=1}^n s_i$.

\paragraph{Ball-end cutter} 
In finishing machining of CNC, a ball-end cutter is commonly employed to finish the surface, modeled as a hemisphere combined with two differently sized cylinders.
\autoref{fig:collision-detection}(a) illustrates the cross-section of the cutter, which can be characterized using four parameters: two radii ($CR$ and $FR$) and two heights ($CH$ and $FH$). To ensure successful manufacturing, it is essential to guarantee the cutter $\mathcal{C}$ does not collide with $M$ in any direction $d_k$:
\begin{equation} 
\forall s_i \in S, \forall d_k \in D, s_i \cap \mathcal{C}(d_k) = \varnothing
\end{equation}

\paragraph{Inaccessible points.} 
For each $s_i \in S$, collision detection is performed with other sites, $\forall s_j\in S$ $(j \neq i)$, using the method from \cite{zhong2023vasco}. We first uniformly sample cutter directions $D = \{d_1, d_2, \dots, d_m\}$ using the Fibonacci Sphere sampling method~\cite{vorobiev2002fibonacci} on the upper Gaussian hemisphere and then rotate $M$ contrarily along each direction. To accelerate computation, a cylinder with a radius of $FR + \sigma$ is added as a detecting box before collision detection, allowing only points within the cylinder to undergo finer detection, as shown in \autoref{fig:collision-detection}(b).
Next, as illustrated in \autoref{fig:collision-detection}(c), each $s_j$ is rapidly evaluated for collision by calculating its horizontal distance from the center of $\mathcal{C}$. If the Z-coordinate of $s_j$ exceeds $CR + CH + FH$, it is immediately classified as colliding with the infinitely large shaft space (called global collision). 
After the traversal, if $s_i$ collides with at least one $s_j$ in all cutter directions, it is classified as an inaccessible point:
\begin{equation} 
s_i \leftarrow l_I \iff \forall d_k \in D, |S \cap \mathcal{C}(d_k)| > 0
\end{equation}
Compared to the triangle-facet-based approach~\cite{dhaliwal2003algorithms}, which involves collision detection between the cutter’s cylindrical surface and the triangular mesh, the proposed method is based on discrete sampling points (sites of Voronoi cells), significantly improving computational efficiency.
Even so, this method still has a worst-case complexity of $O(mn^2)$. Besides, it is crucial to ensure that the shortest edge length of the smallest Voronoi cell is greater than $2*CR$ to prevent the cutter from passing through the cell without detection, as illustrated in \autoref{fig:collision-detection}(d).

\begin{figure}[t]

\centering
\includegraphics[width=1.0\linewidth]{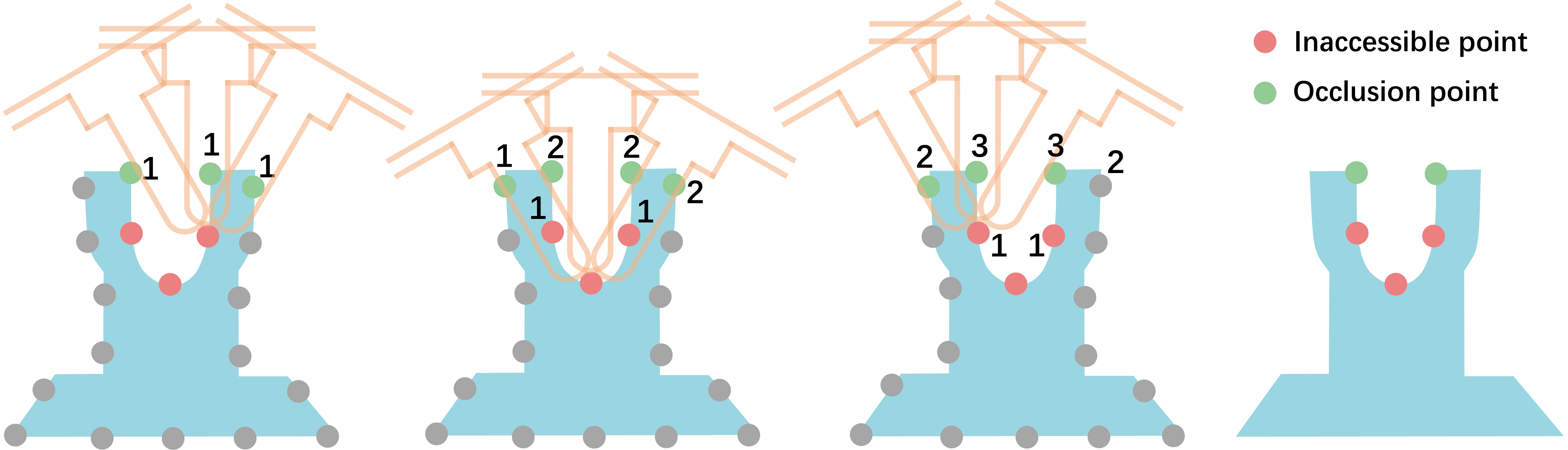}
\leftline{ \footnotesize  \hspace{0.08\linewidth}
            (a)  \hspace{0.225\linewidth}
            (b)  \hspace{0.225\linewidth}
            (c)  \hspace{0.195\linewidth}
            (d)}
\caption{Illustration of occlusion point calculation. (a)$\sim$(c) Perform collision detection for three inaccessible points, recording the points that collide with the cutter in each cutter direction and counting the total number of collisions for each point. (d) The top 10$\%$ of points with the highest total collision counts are labeled as occlusion points ($l_O$).}
  \label{fig:Occlusion-analyze}    
\end{figure}

\subsection{Occlusion analysis}
To further assist designers in modifying the "culprit" causing inaccessible points, we compute the "occlusion factor" $\beta_i$ for each $s_i \in S$ to quantify the severity of its occlusion for the inaccessible points:
\begin{equation} 
\beta_i = \sum_{s_j \in S}\sum_{d_k \in D} \left(
\begin{cases} 
1  \text{, if } s_j \text{ is inaccessible and } s_i \text{ occludes } \text{$s_j$ in} \; d_k \\
0  \text{, otherwise}
\end{cases} \right)
\end{equation}
The top 10$\%$ of $s_i$ with the highest $\beta_i$ values are defined as "occlusion points" ($s_i \leftarrow l_O$). As shown in \autoref{fig:Occlusion-analyze}, the points in the upper region of the 2D shape are marked as "occlusion points."

\section{Results and Discussions}
\label{sec:results}

\subsection{Data processing and datasets}
\paragraph{Data processing and cleaning} 
We performed the accessibility and occlusion analysis described in \autoref{sec:Overview} to construct the training and test sets, uniformly sampled 150 cutter directions on the upper hemisphere. We selected various CAD shapes from the ABC dataset \cite{Koch2019CVPR} and freeform shapes from Thingi10K \cite{Thingi10K}. After cleaning non-manifold, non-watertight, and multi-component assemblies, we retained high-quality shapes. To avoid invalid data from shapes significantly smaller than the cutter, we extracted each shape's bounding box, ensuring its minimum edge length was at least 80 mm. For each shape, the coordinates of Voronoi cell sites, normals, and corresponding inaccessible and occlusion labels were recorded in the training and test sets.

\paragraph{Datasets}
In the training set, we randomly generated the four cutter parameters for over 5K CAD shapes within specified ranges: $CR \in [1, 2]$, $FR \in [5, 100]$, $CH \in [0.1, 10.1]$, and $FH \in [0.1, 10.1]$. Each shape had an average of approximately 7K mesh vertices. 
For the test sets, we selected 1K CAD shapes distinct from the training set and created two sets through remeshing, containing 7K and 15K mesh vertices, respectively. Similarly, we generated two test sets for freeform shapes. 
To further validate DeepMill's generalization ability, we curated a complex dataset of 500 shapes with over 100K vertices from the ABC dataset. Additionally, we varied cutter size parameters and generated multiple test sets based on CAD shapes to demonstrate DeepMill's adaptability to different cutter sizes.

\subsection{Implementation}
All experiments were conducted on a desktop computer equipped with an Intel Core i7-11700F CPU running at 2.5 GHz, 16 GB of memory, and an RTX 3090 GPU with 24 GB of memory. The source code will be released after the paper is published. The octree depth was set to 5, with encoder channels configured as [32, 32, 64, 128, 256] and decoder channels as [256, 256, 128, 96, 96]. The cutter module channels were set to [4, 32, 256]. We used the stochastic gradient descent (SGD) optimizer for training, starting with an initial learning rate of 1.0, which was adjusted using the Cosine Annealing method. The network was trained for 1500 epochs with a batch size of 128 for both training and testing. All input points were normalized to the unit cube $[-1, 1]^3$, and data augmentation methods from~\cite{choy20194d}, including random mirroring and elastic deformations, were applied.

\begin{table}[t]
	\begin{minipage}{1.0\linewidth}
		\centering
		\caption{Statistics of DeepMill and geometric method on various datasets. The values in ( ) indicate the average number of mesh vertices in the dataset. $\text{Acc}_i$ and $\text{Acc}_o$ represent the prediction accuracy for inaccessible points and occlusion points, respectively. $\text{F1}_i$ and $\text{F1}_o$ represent their F1-scores. $T_i$ and $T_o$ denote the calculation time (second) for inaccessible and occlusion points by geometric method, while $T$ represents the total time. Due to the excessive computation time required by geometric method to generate occlusion points for complex models, we do not perform statistics on them.} 
		\label{table:table-different-data}
            \setlength{\tabcolsep}{2pt}
		\resizebox{1\textwidth}{!}{
			\begin{tabular}{l|ccccc|ccc} 
				\hline
				&  \multicolumn{5}{c|}{\textbf{DeepMill}}  & \multicolumn{3}{c}{\small{\textbf{Geometric}}} \\ 
                \hline
                \textbf{DATASET} & $\textbf{Acc}_i$ & $\textbf{F1}_i$ & $\textbf{Acc}_o$ & $\textbf{F1}_o$ &$T$ & $T_i$ & $T_o$ & $T$ \\
                \hline 
                CAD(7K) 	&96.3$\%$ & 97.2$\%$ & 98.3$\%$  & 89.4$\%$ & \cellcolor{orange!25}0.01  & 4.0 & 25.1 & 29.1 \\ 
                CAD(15K) 	& \cellcolor{orange!25}96.3$\%$ & \cellcolor{orange!25}97.3$\%$ & \cellcolor{orange!25}98.3$\%$ & \cellcolor{orange!25}90.0$\%$ & \cellcolor{orange!25}0.01  & 17.3 & 207.4 & 224.7 \\ 
                Freeform(7K) 	& 92.8$\%$ & 93.7$\%$ & 97.5$\%$ & 86.5$\%$ & \cellcolor{orange!25}0.01  & 5.4 & 41.3 & 46.7 \\ 
                Freeform(15K) 	& 93.2$\%$ & 93.3$\%$ & 98.0$\%$ & 88.7$\%$ & \cellcolor{orange!25}0.02  & 19.8 & 373.1 & 392.9 \\ 
                \small{Complex(10W+)}  	& 90.5$\%$ & 90.0$\%$ & $\backslash$ & $\backslash$ & \cellcolor{orange!25}0.04  & 137.0 & $\backslash$ & $\backslash$ \\ 
                \hline
			\end{tabular}
		}
	\end{minipage}
\end{table}

\subsection{Evaluation}
\begin{figure*}[htbp]
\vspace{-5pt}
\centering
  \includegraphics[width=0.99\linewidth]{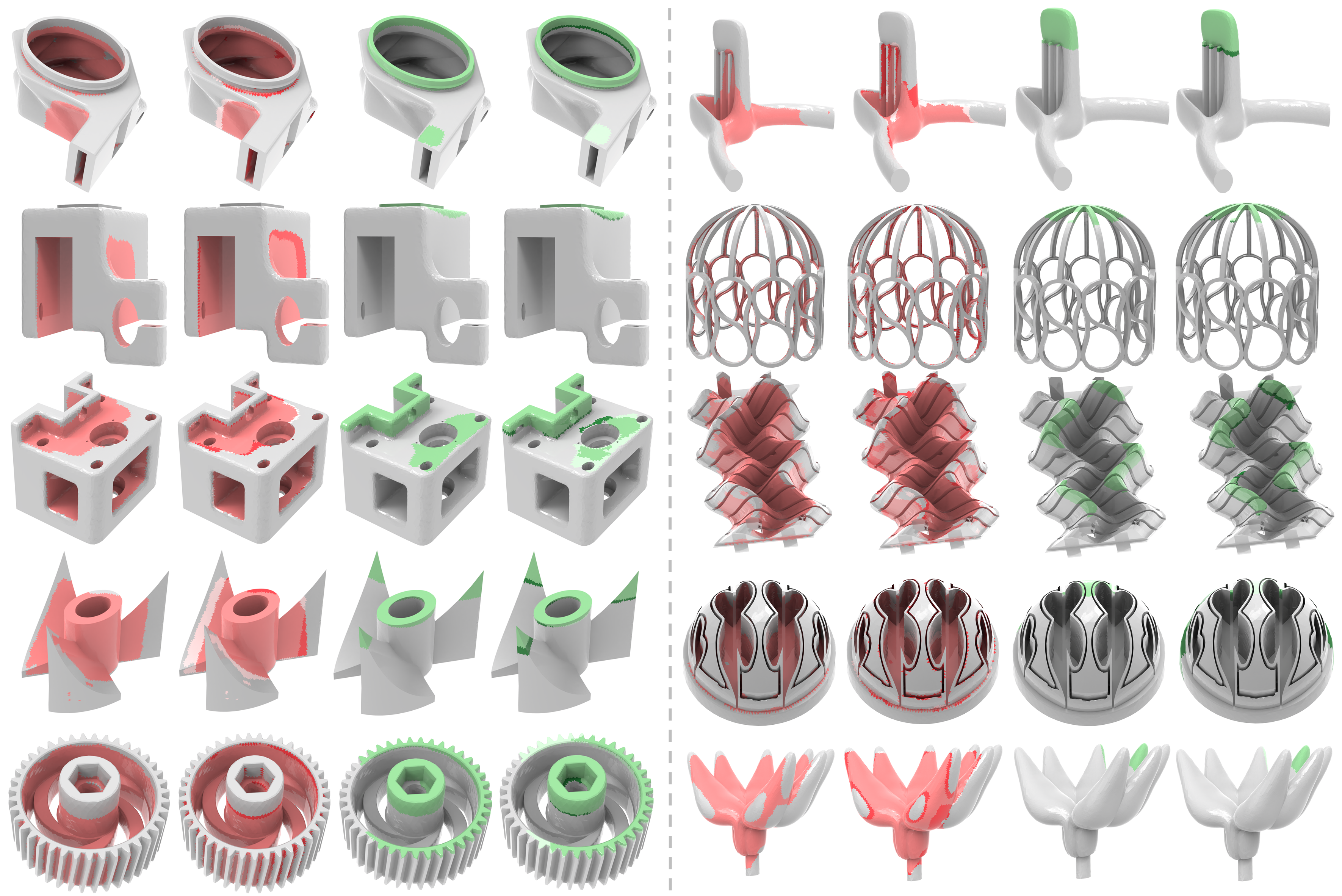}
\vspace{-5pt}
\caption{The gallery of DeepMill prediction results. On the left are CAD shapes, and on the right are freeform shapes. For each row of shapes, the first and third columns show the inaccessible and occlusion regions predicted by DeepMill. In the second and fourth columns, darker shades represent under-predicted areas, while lighter shades indicate over-predicted areas.}
\label{fig:gallery1}    
\end{figure*}

\paragraph{Efficiency and Generalization}
DeepMill was trained on a dataset of over 5K CAD shapes.
We evaluated its efficiency on datasets containing diverse shapes, as shown in \autoref{table:table-different-data}.
We map the Voronoi sites onto the triangles for better visualization.
\autoref{fig:teaser} and \autoref{fig:gallery1} illustrate plenty of examples of inaccessible and occlusion regions predicted by DeepMill with various cutter sizes.

In general, DeepMill maintains high prediction accuracy, with most errors occurring near the boundaries of inaccessible and occlusion regions, which minimally affects the overall distribution.
Additionally, as the mesh resolution increases (e.g., 7K vs. 15K Freeform shapes), finer geometric details further improve DeepMill's prediction accuracy, and the higher resolution amplifies the time advantage of DeepMill over traditional geometric methods.

\begin{figure*}[htbp]
\centering
  \includegraphics[width=0.99\linewidth]{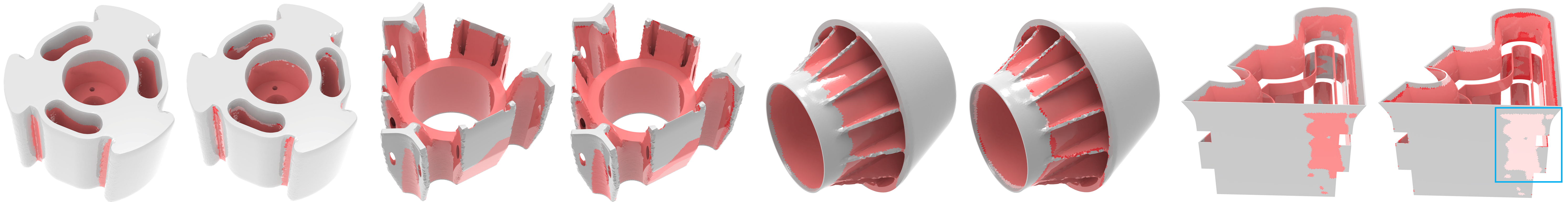}
\vspace{-5pt}
\caption{Testing of DeepMill on complex shapes. For each mesh, the left side shows the prediction results from DeepMill, while the right side displays the differences compared to the geometric method.}
\label{fig:Complex-models}    
\end{figure*}

For different datasets, DeepMill exhibits varying performance:
1) On CAD datasets with geometric styles similar to the training set, DeepMill achieves up to 96.3$\%$ accuracy for predicting inaccessible regions. Since occlusion regions are derived from inaccessible regions, they present greater prediction challenges.
2) For freeform shapes, which feature fewer sharp or weak geometric characteristics and differ significantly from CAD shapes, the network maintains high prediction accuracy (92.8$\%$, 86.5$\%$), demonstrating its ability to effectively learn the geometric relationship between shapes and cutters.
3) The Complex dataset consists of meshes with a large number of triangular facets (over 100K), as shown in \autoref{fig:Complex-models}. The intricate geometries increase prediction difficulty, and for unconventional structures that diverge significantly from the training set, DeepMill show reduced accuracy (e.g., top-right corner of \autoref{fig:Complex-models}).

\paragraph{Computation time}
DeepMill offers significant advantages over traditional geometric methods, achieving real-time predictions. For inaccessible and occlusion regions, DeepMill requires only \textbf{0.004$\%$} of the total time needed by geometric methods on CAD shapes with 15K mesh vertices. For more complex shapes, the time is reduced to \textbf{0.029$\%$} for inaccessible analysis only. Moreover, if finer cutter direction sampling is used in geometric methods, the time efficiency of DeepMill becomes even more pronounced.

\subsection{Ablation and Comparisons}
\begin{table}[t]
	\begin{minipage}{1.0\linewidth}
		\centering

		\caption{Comparison statistics of the cutter module in DeepMill for test sets with different cutter size. The baseline refers to the control group without the cutter module. $Avg_i$ and $Avg_f$ represent the average values of accuracy and F1-score, respectively. In the Uniform test set, the random range of cutter parameters is the same as in the training set. In the Short set, the ranges of $CH$, $FR$, and $FH$ are [0.1, 0.2], [80, 100], and [0.1, 0.2], respectively. In the Long set, the ranges are [10, 10.1], [5, 5.1], and [10, 10.1]. In the Extreme set, the ranges are [20, 20.1], [5, 5.1], and [20, 20.1].}
		\label{table:table-time}
            \setlength{\tabcolsep}{3pt}
		\resizebox{1\textwidth}{!}{
			\begin{tabular}{l|l|ccccccc}
				\hline
				\textbf{Cutter} & \textbf{Decoder}   &  \textbf{$\text{Acc}_i$} & \textbf{$\text{F1}_i$} & \textbf{$\text{Acc}_o$} & \textbf{$\text{F1}_o$} & \textbf{$\text{Avg}_{acc}$} & \textbf{$\text{Avg}_{f1}$} \\ \hline 
                
                \multirow{2}{*}{Uniform}
                &Baseline 	& 0.932 & 0.949 & 0.976  & 0.856 & 0.954 & 0.903\\ 
                &Our	& \cellcolor{orange!25}0.963 & 0.972 & 0.983 & \cellcolor{orange!25}0.894 & 0.973 & 0.933\\
                \hline
                \multirow{2}{*}{Short}
                &Baseline 	& 0.898 & 0.916 & 0.972 & 0.838 & 0.935 & 0.877\\ 
                &Our	& \cellcolor{orange!25}0.962 & 0.967 & 0.981 & \cellcolor{orange!25}0.896 & 0.972 & 0.932\\ 
                \hline
                \multirow{2}{*}{Long}
                &Baseline 	& 0.869 & 0.909 & 0.930 & 0.576 & 0.900 & 0.743\\ 
                &Our 	& \cellcolor{orange!25}0.940 & 0.960 & 0.961 & \cellcolor{orange!25}0.763 & 0.951 & 0.862\\ 
                \hline
                \multirow{2}{*}{Extreme}
                &Baseline & 0.809 & 0.873 &  0.901 & 0.388 & 0.855 & 0.631\\ 
                &Our	& \cellcolor{orange!25}0.928 & 0.937 & 0.975 & \cellcolor{orange!25}0.537 &  0.952 & 0.737\\ 
                \hline
			\end{tabular}
		}
	\end{minipage}
\end{table}

\begin{figure*}[htbp]
\centering
  \includegraphics[width=0.99\linewidth]{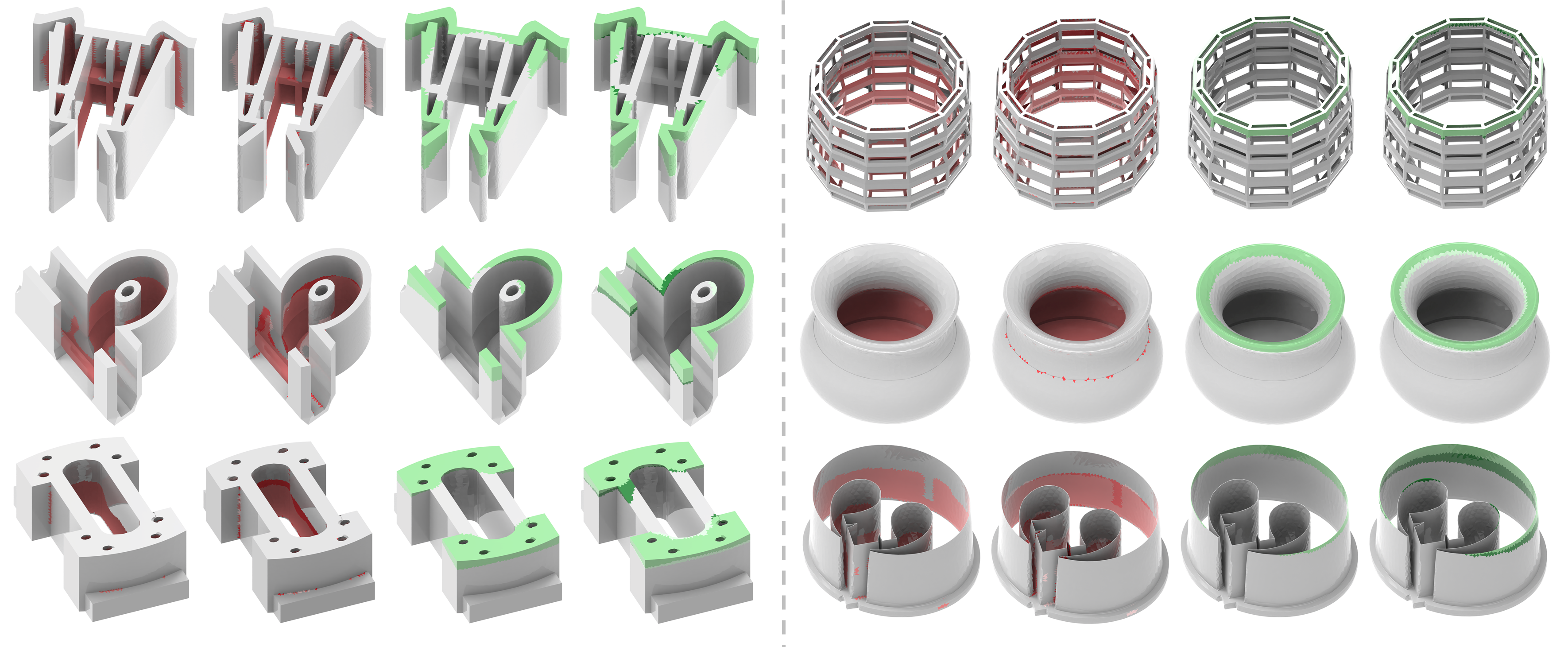}
\vspace{-5pt}
\caption{Demonstration of DeepMill prediction results with extreme size of cutter. After adding the dataset generated with extreme cutters to the training set, DeepMill was able to extrapolate its prediction capability to cases involving extreme cutters.}
\label{fig:gallery2}    
\end{figure*}

\paragraph{Ablation study for cutter module}
To demonstrate the effectiveness of the cutter module, we compared it with the baseline model without the cutter module. As shown in \autoref{table:table-time}, DeepMill with the cutter module significantly outperforms the baseline across test sets with various cutter parameter ranges. 
This improvement is especially noticeable when using short and long cutters, as the baseline model tends to predict using an "average-sized cutter," resulting in a larger accuracy gap compared to DeepMill.
It is worth noting that some regions of the shape are inherently "cutter-independent," meaning their accessibility remains unchanged regardless of cutter size. The baseline successfully learns these regions, maintaining reasonable prediction accuracy. 
On the other hand, since shorter cutter sizes were used in the training set, the ratio of local collisions to global collisions is relatively small. As a result, DeepMill performs less accurately in predicting occlusion regions caused by extensive local collisions from extremely long cutters.
To demonstrate it, we added 3K+ CAD shapes with extreme cutter sizes to the training set for calculating inaccessible and occlusion regions. The accuracy and F1-score for the two regions improved to $95.8\%$ and $87.8\%$, respectively. \autoref{fig:gallery2} shows several examples.

\begin{figure*}[htbp]
\centering
  \includegraphics[width=0.99\linewidth]{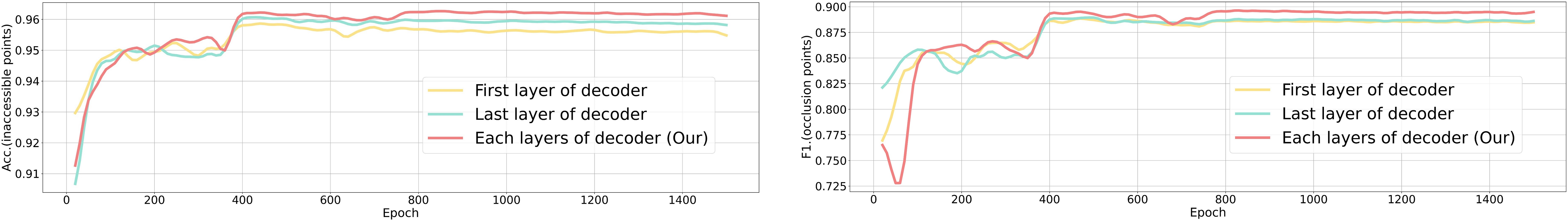}
\vspace{-5pt}
\caption{Comparison of cutter module concatenation methods. The left and right show the prediction accuracy of inaccessible points and the F1 score of occlusion regions for different concatenation methods on the same test set. Our approach performs the best in both measures.}
\label{fig:Comparison-cutter-module}    
\end{figure*}

\paragraph{Comparison of cutter module positions}
We compared the impact of adding the cutter module at different positions in the decoder on the prediction results. As shown in \autoref{fig:Comparison-cutter-module}, adding the module to each layer, rather than only to the first or last layer of the decoder, enables the network to better learn the influence of cutter parameters on both local and global geometry.

\begin{figure}[htbp]
\vspace{-5pt}
\centering
  \includegraphics[width=1.0\linewidth]{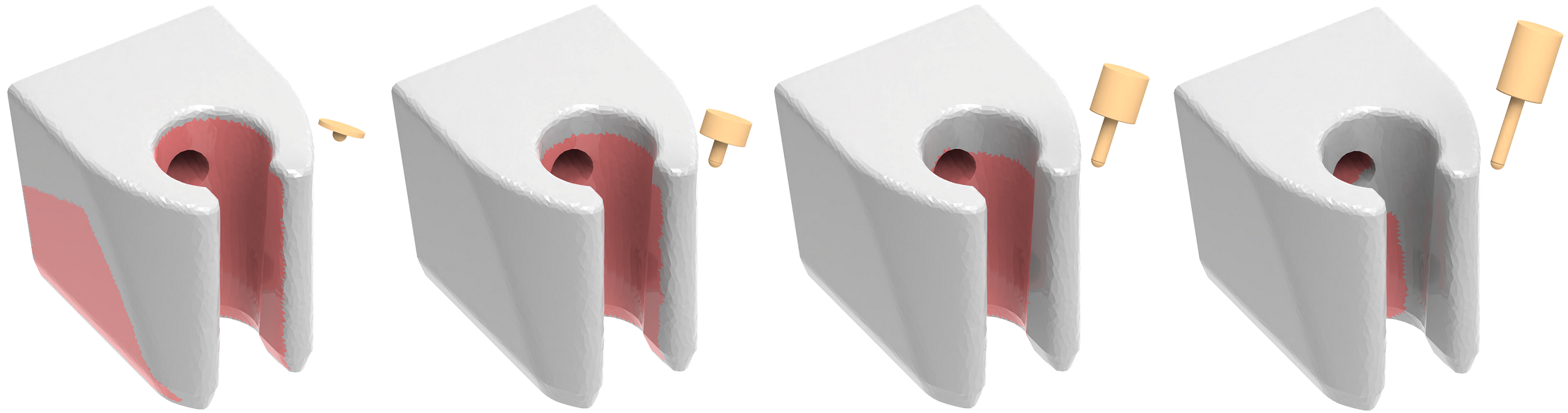}
\vspace{-20pt}
\caption{Illustration of the effect of cutter length on inaccessible regions. Generally, longer cutters lead to fewer inaccessible regions.}
\label{fig:different-cutter}    
\end{figure}

\paragraph{Comparison of different cutter sizes}
\autoref{fig:different-cutter} illustrates the effect of cutter length on inaccessible regions. Short cutter is more prone to collide with the shaft above cutter, resulting in larger inaccessible regions. On the other hand, longer cutter is less likely to cause collisions, allowing for more accessible regions.

\begin{figure}[htbp]
\vspace{-5pt}
\centering
  \includegraphics[width=1.0\linewidth]{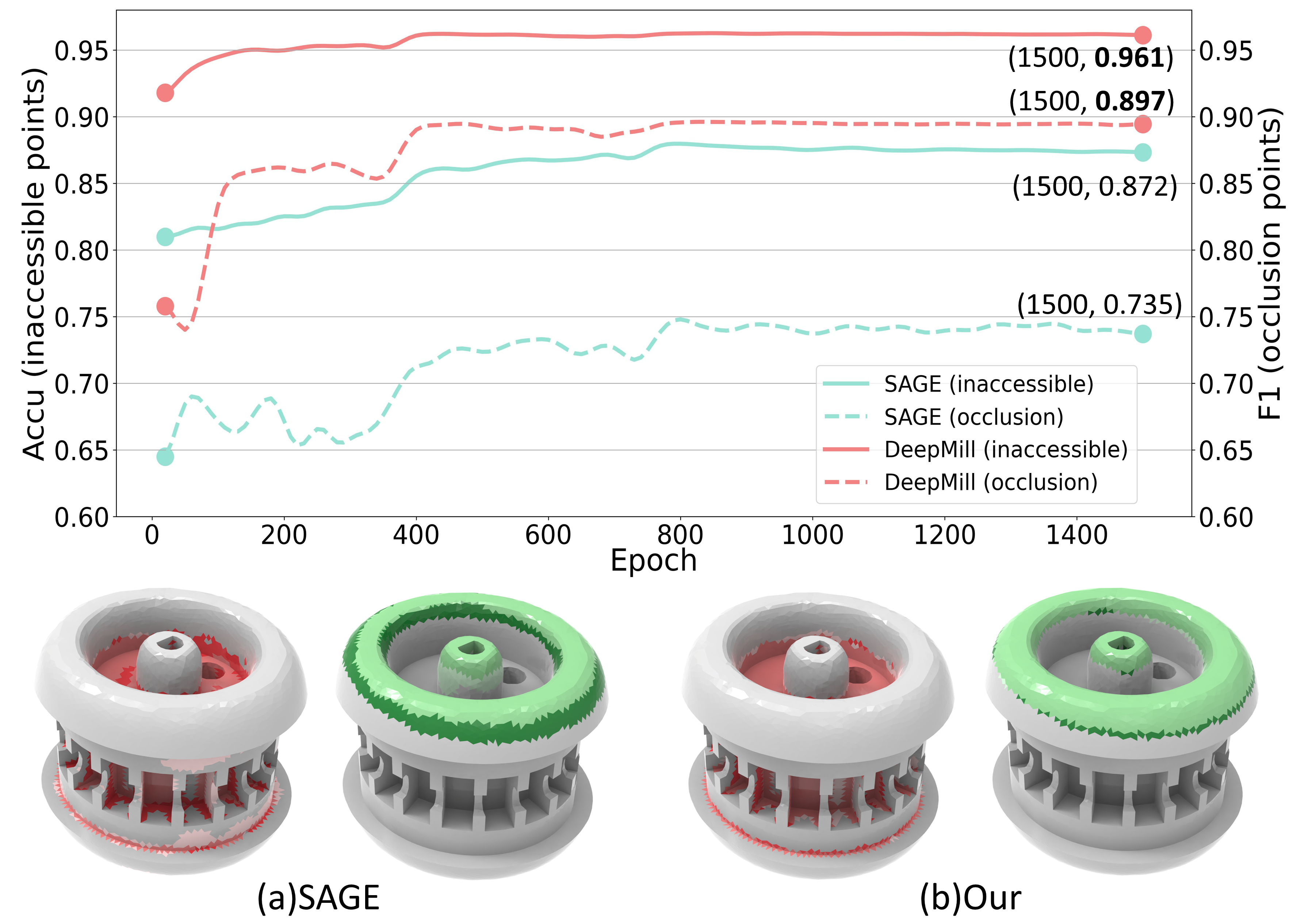}
\vspace{-20pt}
\caption{Comparison with SAGE. DeepMill shows significantly better prediction capabilities for inaccessible and occlusion regions compared to SAGE.}
\label{fig:Ablation-study}    
\end{figure}

\paragraph{Comparison with other network}
\autoref{fig:Ablation-study} presents a comparison between DeepMill and the GraphSAGE baseline. DeepMill significantly outperforms GraphSAGE, demonstrating its superior ability to capture complex geometric interactions. In contrast, GraphSAGE struggles to achieve satisfactory accuracy improvements, highlighting its limited capability to effectively learn collision relationships between shapes and cutters.

\subsection{Discussion and Extension}

\begin{figure}[htbp]
\vspace{-5pt}
\centering
  \includegraphics[width=1.0\linewidth]{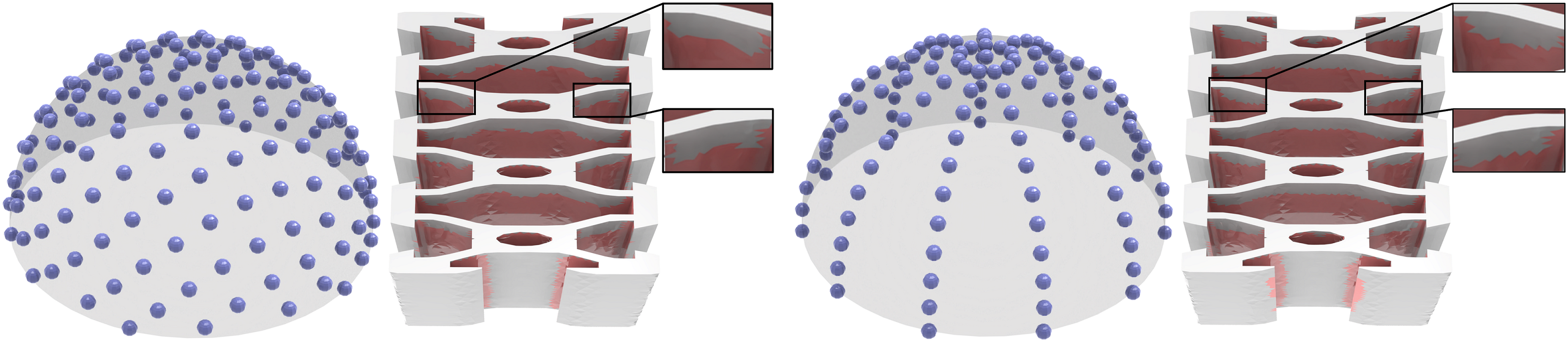}
\leftline{ \footnotesize  \hspace{0.11\linewidth}
            (a)  \hspace{0.18\linewidth}
            (b)   \hspace{0.205\linewidth}
            (c)  \hspace{0.18\linewidth}
            (d)  }
\vspace{-15pt}
\caption{Geometric symmetry illustration. (a) Non-axisymmetric cutter sampling causes asymmetric inaccessible regions (b). (c) Axisymmetric method has uneven distribution. (d) DeepMill combines both, yielding more symmetrical inaccessible regions.}
\label{fig:symmetry}    
\end{figure}

\paragraph{Geometric symmetry}
The geometric method in the dataset uses Fibonacci sphere sampling for evenly distributed cutter directions, as shown in \autoref{fig:symmetry} (a). However, the directions lack axial symmetry, causing asymmetric inaccessibility distributions for symmetric shapes, as shown in \autoref{fig:symmetry} (b).
Another sampling method based on spherical coordinates, shown in \autoref{fig:symmetry} (c), achieves symmetric direction distribution but suffers from uneven spacing, easily missing directions.
Surprisingly, DeepMill combines the strengths of both methods, learning to symmetrically adjust predictions from uniformly distributed directions. It produces more symmetric and reasonable inaccessibility regions when predicting geometrically symmetric shapes, as illustrated in \autoref{fig:symmetry} (d).

\begin{figure}[htbp]
\centering
  \includegraphics[width=1.0\linewidth]{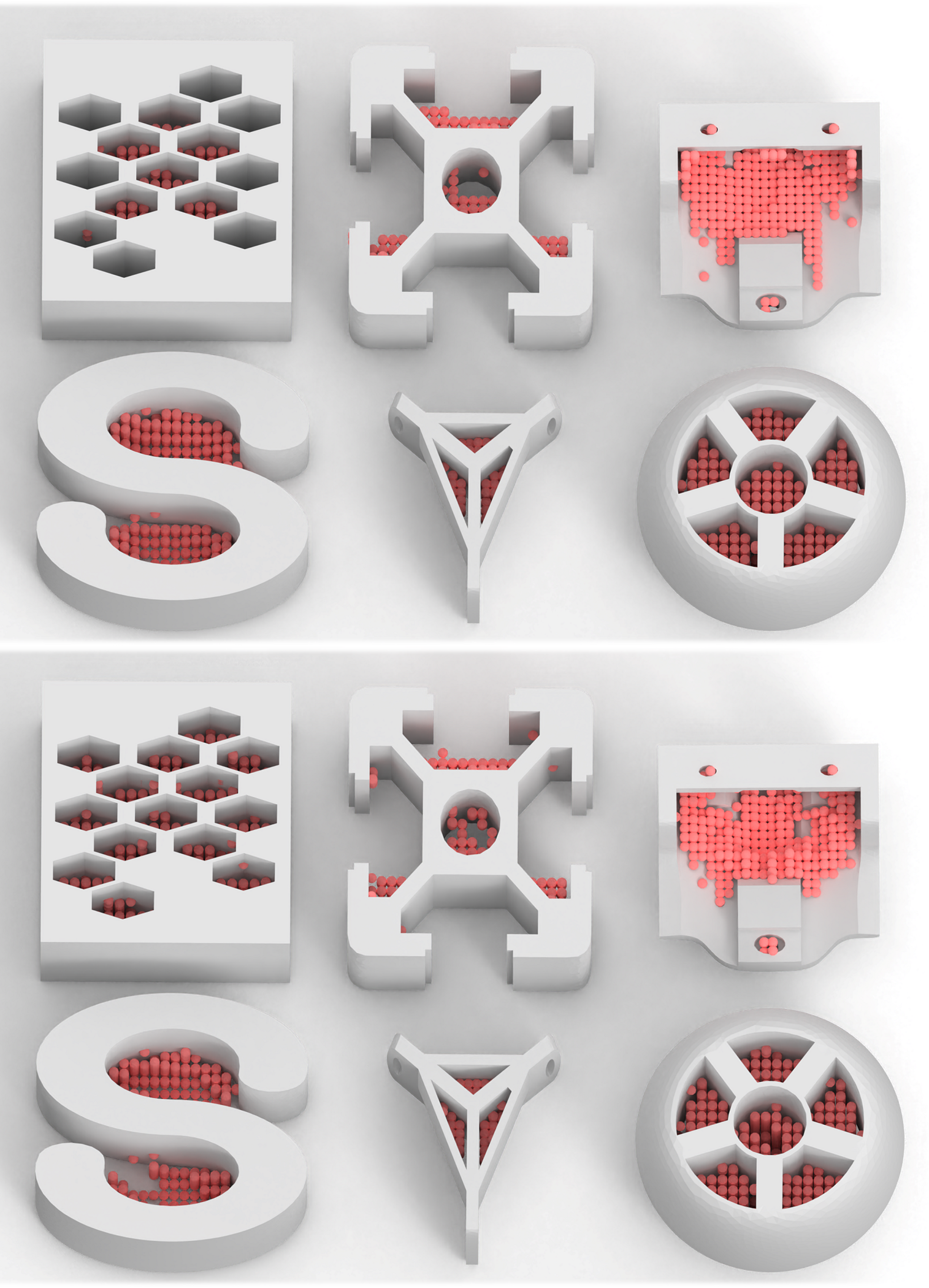}
\vspace{-7pt}
\caption{Illustration of accessibility analysis within the volume. The red points represent inaccessible sampling points. On the top, the results predicted by DeepMill are shown, and on the bottom, the results obtained by the geometric method are displayed.}
\label{fig:volume-accessibility}    
\end{figure}

\paragraph{Volume accessibility analysis}
DeepMill can also be applied in accessibility analysis for other machining methods. For instance, rough machining removes the blank material layer by layer using a mill cutter, which is typically the pre-process for the finishing process. The accessibility analysis focuses on the interior volume of the blank. 
We use the same collision detection method to detect the interior sampling points of the bounding box of the input mesh to generate datasets. As shown in \autoref{fig:volume-accessibility}, after training on the new dataset of over 5K CAD shapes, DeepMill can accurately predict the inaccessible regions within the volume.
Compared to surface-based accessibility analysis, DeepMill can more easily predict accessibility within the volume, achieving an accuracy of up to $97.9\%$. Volume sampling uses voxelization, where adjacency resembles pixels in 2D, making it more suitable for 3D convolution operations.

\section{Conclusion and Future work}
\label{sec:conclusion}
This paper introduces DeepMill, a deep learning framework that improves cutter accessibility and manufacturability analysis for complex designs. Utilizing octree-based convolutional neural network (O-CNN), DeepMill efficiently predicts inaccessible regions and occlusions across various cutter sizes, overcoming the scalability and computational limitations of traditional methods. Its real-time predictions enable faster design iterations and enhance production efficiency. Additionally, the new dataset introduced supports further research and development of robust manufacturability analysis cutters, making DeepMill a significant advancement in the field. 
Extensive testing and comparisons have demonstrated DeepMill's powerful cutter-aware prediction ability.




Based on the challenges and opportunities outlined, several directions for future work are proposed. First, to enhance the prediction performance of our current network, integrating an attention mechanism will be a promising approach. 
Second, incorporating geometric prior knowledge, such as symmetry, similarity, and topological properties, will allow the network to better capture the underlying structures.
%
Another promising avenue involves incorporating cutters that work with irregular shapes in subtractive manufacturing, which will broaden the applicability of our framework to a wider range of real-world scenarios.
Additionally, exploring the downstream applications of our current framework, such as path planning and model correction from non-accessible to accessible models, would be an exciting direction for further development.

Exploring learning-based methods in digital design and manufacturing, integrating both subtractive and additive aspects, will drive intelligent solutions to complex challenges.



\bibliographystyle{formats/ACM-Reference-Format}
\bibliography{neural} 

\clearpage

\end{document}